\documentstyle[aas2pp4]{article}

\def\RXTE{{\it RXTE}}

\lefthead{DELGADO-MARTI et al.}
\righthead{The Orbit of X Per}

\begin{document}

\title{The Orbit of X Per and Its Neutron Star Companion}
\author{Hugo Delgado-Marti, Alan M. Levine, Eric Pfahl, \& Saul A. Rappaport}

\affil{
Department of Physics and Center for Space Research,
Massachusetts Institute of Technology, Cambridge, MA 02139; aml@space.mit.edu
}

\authoraddr{Center for Space Research, M.I.T., Room 37-575,
   Cambridge, MA 02139-4307}
\authoremail{aml@space.mit.edu}

\begin{abstract}

We have observed the Be/X-ray pulsar binary system X Per/4U~0352+30 on
61 occasions spanning an interval of 600 days with the PCA instrument
on board the {\it Rossi X-ray Timing Explorer} (\RXTE\,).  Pulse
timing analyses of the 837-s pulsations yield strong evidence for the
presence of orbital Doppler delays.  We confirm the Doppler delays by
using measurements made with the All-Sky Monitor on \RXTE\,.  We infer
that the orbit is characterized by a period $P_{orb} = 250$ d, a
projected semimajor axis of the neutron star $a_x \sin i = 454$ lt-s,
a mass function $f(M) = 1.61$ M$_{\sun}$, and a modest eccentricity $e
= 0.11$.  The measured orbital parameters, together with the known
properties of the classical Be star X Per, imply a semimajor axis $a =
2.2$ AU, and an orbital inclination $i \sim 23\arcdeg$--$30 \arcdeg$.

We discuss the formation of the system in the context of the standard
evolutionary scenario for Be/X-ray binaries.  The orbital eccentricity
just after the supernova explosion was almost certainly virtually the
same as at present, because the Be star is much smaller than the
orbital separation.  We find that the system most likely formed from a
pair of massive progenitor stars, and probably involved a quasi-stable
and nearly conservative transfer of mass from the primary to the
secondary.  We find that the He star remnant of the primary most
likely had a mass $\lesssim\,6$ M$_{\sun}$ after mass transfer.  If
the supernova explosion was completely symmetric, then the present
orbital eccentricity indicates that $\lesssim$\,4 $M_{\sun}$ was
ejected from the binary.  If, on the other hand, the birth of the
neutron star was accompanied by a ``kick'' of the type often inferred
from the velocity distribution of isolated radio pulsars, then the
resultant orbital eccentricity would likely have been substantially
larger than 0.11.  We have carried out a Monte Carlo study of the
effects of such natal kicks, and find that there is less than a 1$\%$
probability of a system like that of X Per forming with an orbital
eccentricity $e \lesssim 0.11$.  Finally, we speculate that there may
be a substantial population of neutron stars formed with little or no
kick.

\end{abstract}


\section{INTRODUCTION}

X Per is a bright and highly variable star with a visual magnitude
that ranges from $\sim$\,6.1 to $\sim$\,6.8 (\cite{moo74,roc97}).  When
the star is particularly bright, its spectrum displays strong emission
in H$\alpha$ and other Balmer lines which marks this clearly as a
Be-type system; when the star is faint, the emission lines disappear
and it appears to be a normal early-type star (\cite{fab92,roc97} and
references therein).  The variability is commonly supposed to be
caused by the formation and dissipation of a disk around the star, as
in other Be stars, and appears to require free-free and free-bound
emission within the disk in addition to electron scattering of the
photospheric flux (e.g., \cite{kun95,tel98}).  The spectral class of
the underlying OB star has been estimated to be O9.5\,III to B0\,V
(\cite{sle82,fab92,lyu97}).  Based on spectroscopic parallax, distance
estimates range from $700 \pm 300$ pc up to $1.3 \pm 0.4$ kpc
(\cite{fab92,lyu97,roc97,tel98}).  Lyubimkov et al. (1997) have used
both spectroscopic and photometric observations of the low-luminosity
diskless phase of X Per to infer the characteristics of the visible
component; their results indicate that X Per is likely to have a mass
of $\sim$\,13-20 M$_{\sun}$ and a radius of 5-10 R$_{\sun}$.

X Per is also the optical counterpart of the low-luminosity X-ray
source 4U~0352+30 (\cite{bm72,vdb72}).  X-ray observations have,
furthermore, revealed pulsations with a period of $\sim$\,835 s
(\cite{whi76}).  This indicates that X Per must be a binary system
containing both a Be star and either a slowly rotating neutron star or
a white dwarf. For the first 5 years after the discovery of
pulsations, the pulsar exhibited apparently erratic pulse frequency
variations superposed on a long-term trend in which it was spinning up
at the rate of $\dot{P}_{pulse}/P_{pulse} = -1.5 \times 10^{-4}$
yr$^{-1}$.  This was followed by $\sim$\,20 years of spindown at a
similar absolute rate (\cite{whi76,rob96,dis98} and references
therein).  In all, there have been only about two dozen determinations
of the pulse period of 4U~0352+30, so that its pulse period behavior
on short timescales (e.g., weeks, months, and even a year) is not well
documented.  The X-ray luminosity varies on long timescales (years)
from as high as $\sim 3 \times 10^{35}$ ergs s$^{-1}$ to as low as its
current value of $\sim 5 \times 10^{34}$ ergs s$^{-1}$ (for an assumed
distance of 1.3 kpc; \cite{roc93}; also see \cite{dis98}).

The observed pulse period variations strongly suggest that the X-ray
source is an accreting neutron star, with the period variations
largely due to accretion torques (see, e.g., \cite{bil97} and
references therein).  White et al. (1982) have argued that
considerations of the X-ray luminosity and spectrum in the context of
accretion from a stellar wind strongly indicate that 4U~0352+30 is a
neutron star.  Haberl et al. (1998) have noted that the 837 second
rotation period is one of the longest known for any neutron star.

The only report of an orbital period for X Per ($\sim$\,580 days) was
given by Hutchings et al. (1974) and was based on optical spectroscopy
(also see \cite{hut75,hut77}).  Penrod \& Vogt (1985) explained the
radial velocity results reported by these authors by emission variably
filling in the Balmer absorption lines, so that the apparent radial
velocity curve did not represent the velocity of X Per.  This cast
doubt on the interpretation of $\sim$\,580 days as the orbital period.
Moreover, this period has never been securely confirmed by any
subsequent optical or X-ray observations
(\cite{wei84,rey92,roc93,smi99}).  There have been other suggestions
of orbital periods based on weak evidence.  In retrospect, one of the
most interesting of these was a possible $\sim$\,250 d periodicity in
pulse period measurements noticed by White et al. (1982).

There is one line of evidence that qualitatively supports the idea of
a long orbital period for X Per, viz. the ``Corbet diagram''
(\cite{cor86,vdh87}) in which the pulse period is plotted versus
orbital period for the known accretion powered X-ray pulsars.  For the
$\sim$\,10 systems containing Be stars there is a good correlation
between pulse period and orbital period which can be represented by
the relation
\begin{equation}
P_{orb}  \sim  18 (P_{pulse}/1\,\rm{s})^{2/5}\  \rm{days}
\end{equation}
with a residual scatter of about a factor of two.  For a pulse period
of 837 s, eq. (1) predicts an orbital period of $\sim$\,266 d.

We have observed 4U~0352+30 with the Proportional Counter Array (PCA)
on the {\it Rossi X-ray Timing Explorer} (\RXTE\,) in order to track
the pulse phase of the neutron star over a substantial interval of
time and to thereby allow us to search for a Doppler signature of
orbital motion.  We have also utilized observations of 4U~0352+30 made
with the \RXTE\ All-Sky Monitor (ASM) in our search.  In Section 2 we
describe the \RXTE\ observations, and provide a log of observation
dates, durations, and pulse arrival times.  The pulse timing analysis
of the PCA data and the results which indicate an orbit with a period
of 250 days are described in Section 3.  In Section 4 we examine the
possibility that the observed pulse arrival time delays arose from a
random walk in the pulse phase due to accretion torques, and conclude
that this is quite unlikely.  In this section, we also present the
results of our successful search for orbital Doppler delays in
observations made with the ASM.  The astrophysical implications of a
wide yet only mildly eccentric orbit in a Be star X-ray binary are
discussed in the context of binary evolution scenarios and natal
neutron star kicks in Section 5.

\section{OBSERVATIONS}

Observations of X Per (4U~0352+30) have been carried out with the PCA
and HEXTE instruments on \RXTE\ over an interval of 600 days.  Each
observation typically spans about 7000 seconds ($\sim$\,8.5 pulse
periods), with a $\sim$\,2000 sec gap due to Earth occultations.  The
first few observations were performed at approximately 2 day
intervals.  The interval between later observations was gradually
increased to about 2 weeks after one year, with the exception of a
$\sim$\,62 d interval with no observations when the Sun passed through
the nearby region of the sky.  The observations continued past the
first year with a nearly constant spacing of about 2 weeks.  Most
recently, data were acquired at intervals of three weeks.  A log of
the observations is given in Table 1.  The data acquired from the PCA
were telemetered in ``Goodxenon1'' and ``Goodxenon2'' modes which
preserve the 1 microsecond time resolution and the inherent energy
resolution of the detectors.\placetable{table1}

The count rates from 4 of the 61 PCA observations of 4U~0352+30 are
shown in Figure 1.  The individual 837-s pulsations are clearly
evident, although there is significant structure from pulse to pulse
as well as on shorter timescales.  The variability exhibited in Figure
1 is typical of all the observations.

The All-Sky Monitor (\cite{lev96}) onboard \RXTE\ has obtained some
21,000 intensity measurements of 4U 0352+30 over the past 4 years.
Each measurement is an average over an interval of 90 s, which is
sufficiently short to allow the possibility of observing the 837-s
pulsations.  The mean count rate in the ASM of 4U~0352+30 is 0.69
s$^{-1}$, and the mean value of the signal to noise ratio of
individual observations is only $\sim\,0.7$.  Nonetheless, the large
number of observations allows meaningful results to be obtained.

Background-subtracted count rates averaged over time intervals of
$\sim$\,5000 s or more, as seen in both the PCA and ASM, are shown in
Figure 2 for the entire 4 year interval covered by the observations.
There are no obvious trends or periodic variations in the source
intensity as seen by either instrument.  A quantitative analysis of
temporal variability using these intensity data is presented in Section
4.3.

\section{ANALYSIS AND RESULTS}

The 2--20 keV band count rate data from each of the PCA observations
were folded modulo a trial pulse period, after the observation times
were adjusted to the Solar System barycenter. Each folded pulse
profile was then cross correlated with a pulse template to find a
pulse arrival time for that observation.  In practice, this procedure
was iterated to improve the pulse template by averaging phase-aligned
profiles from individual observations.  Details of this procedure can
be found in many references (e.g., \cite{lev00}).  The 61 barycentric
pulse arrival times are given in Table 1.

Background-subtracted pulse profiles in several energy bands are shown
in Figure 3; the 2--20 keV profile was used for the template.  The
pulse profile has a shape that is somewhere between triangular and
sinusoidal, with a modulation fraction of about 50\% (peak-to-peak
amplitude divided by the mean).  We have found that the basic shape of
the pulse profile did not vary over the course of the PCA
observations, nor did it depend significantly on energy band within
the range 2--20 keV.

In Figure 4 we show the results of making different assumptions about
the number of pulses that occurred between each of our measured
arrival times.  For each sequential pair of pulse arrival times, we
divided the time interval by the number of pulses, $n$, we believe
occurred between the two arrival times to obtain an average
pulse period for the interval.  We also show the pulse periods that
would be deduced if the cycle count had been $n+3$, $n+2$, $n+1$,
$n-1$, $n-2$, or $n-3$.  In any plausible accretion torque scenario
for this neutron star (see the discussion below), the pulse period
cannot have changed by more than a fraction of a second over an
interval of 600 days.  Therefore, the plot shows that the pulse period
must be close to 837.5 s and that the pulse count between observations
is generally unambiguously determined.  The only interval in which the
correct pulse count is not obvious from Figure 4 is the long gap in
the observations between day 298 and day 360.  Below, we discuss
evidence obtained from a detailed coherent analysis of the pulse arrival
times that indicates that we have correctly determined the pulse count
in this long interval.  The pulse numbers that we believe to be
correct are listed in Table 1.

Once pulse numbers have been assigned to the arrival times, we can
subtract off the best fitting constant pulse period, and examine the
residual pulse arrival time delays.  The result is shown in the top
panel of Figure 5.  There are clear delays as large as approximately
plus and minus a full pulse period.  We next fit a model including
constant, linear, and quadratic terms to the pulse arrival times to
simulate a simple spin up or spin down of the neutron star.  The
residuals with respect to this simple model are shown in the lower
panel of Figure 5; they suggest orbital motion with a period of
$\sim$\,250 days.

We next carried out a formal search for best-fitting Keplerian orbits.
Models of both circular and mildly eccentric orbits were fit to the
arrival times.  The circular orbit fits included 5 parameters to be
determined, i.e., constant, linear, and quadratic terms to model the
pulsar spin and terms proportional to $\sin \omega_{orb}t$ and $\cos
\omega_{orb}t$ to account for orbital Doppler delays.  In searching
for eccentric orbit solutions, we utilized the approximation wherein
small eccentricities produce sinusoidal modulations in the pulse
arrival times at twice the orbital frequency.  The small orbital
eccentricities found (see the discussion below) justify this
approximation.  Therefore, our eccentric orbit fits included terms
proportional to $\sin 2\omega_{orb}t$ and $\cos 2\omega_{orb}t$ as
well as the 5 terms used in the circular orbit fits.  The circular and
eccentric orbit fits were each repeated for closely spaced values of
the orbital period in the range 30 to 600 days.

A plot of the magnitude of the root-mean-square (rms) pulse arrival
time fit residuals vs. orbital period for circular and mildly
eccentric orbits is shown in Figure 6. This figure shows that there
are clearly defined best-fit solutions at orbital periods very close
to 250 days.

The pulse arrival time delays are shown together with the best-fit
circular and eccentric orbit models in Figure 7.  For this figure the
constant, linear, and quadratic terms of each fit have been subtracted
from both the arrival times and the models.  The circular orbit has a
period of 249.9 days, while the eccentric orbit fit yields a period of
250.3 days, with a corresponding eccentricity of $0.111 \pm 0.018\ (1
\sigma$ confidence).  The orbital parameters are summarized in Table
2.  The statistical uncertainties in the orbital parameters were
determined by using the rms residuals from the best fit as a measure
of the uncertainty in the individual pulse arrival times.  The
eccentric orbit fit yields an rms scatter about the fit of 21.8 s,
compared with 27.7 s for the circular orbit fit.  We interpret this
difference as being significant at the $\sim\,6 \sigma$ level. We
found that the effects of any correlation of the other parameters with
the orbital period were small; the $1\sigma$ error estimates in Table
2 did not need to be adjusted for this effect.  \placetable{table2}

Earlier we noted that determination of the pulse count during the long
gap in the observations required a more detailed analysis.  We have
therefore carried out orbital fits under the assumption that the
number of pulses in the gap was one more or less than we assumed for
the analysis given above.  Neither case yielded a fit with an rms
value less than 79 s for any orbital period in the range 30 to 350 d
(see Fig. 6).  Both cases could thus be ruled out with high
confidence.

The value of the pulse period derivative that we derive from the best
fitting circular and eccentric orbits is $\dot{P}_{pulse}/P_{pulse} =
1.26 \pm 0.01 \times 10^{-4}$ yr$^{-1}$ (see Table 2).  The positive
derivative indicates that the neutron star is spinning {\it down}.
This is consistent with the long-term pulse period changes observed in
X Per over the past $\sim$\,20 years.

The orbital parameters discussed above for X Per yield a mass function
of $1.61 \pm 0.05$ M$_{\sun}$ (see Table 2).  If we assume that the
mass of X Per is in the range of 15--30 M$_{\sun}$, and that the
neutron star has a mass of 1.4 M$_{\sun}$, then the orbital
inclination lies in the range of $23\arcdeg$--$30\arcdeg$.  In turn,
this implies that the semimajor axis of the orbit is about 2.2 AU in
size.

\section{ATTEMPTS TO VERIFY THE ORBIT}

\subsection{Simulations with Random Walk in Pulse Phase}

Since a number of X-ray pulsars exhibit spinup as well as spindown
episodes on a wide range of timescales (e.g., Bildsten et al. 1997),
it is worth asking whether the sinusoidal component of the variation
of the pulse arrival time delays of 4U\,0352+30 could be due to
variations in the intrinsic pulsar spin rate instead of orbital
motion.  We therefore carried out simulations of a model of intrinsic
spin rate variations to test this alternate hypothesis.  The model we
selected, that of random white torque noise, while hardly unique, can
at least test the plausibility that the observed pulse arrival time
delays could be generated by torques on the neutron star (see, e.g.,
\cite{dee89}).  For these simulations, we generated artificial sets of
pulse arrival times corresponding to a set of observations covering a
600-day interval, and spaced in a manner roughly approximating that of
the \RXTE\ observations.  For each simulated data set, a string of
6000 Gaussian-distributed random numbers was generated to represent
the values of the torque averaged over 0.1 day time intervals.  The
amplitude of the Gaussian random noise was set so as to produce, after
many such simulations, an $rms$ value of $\dot{P}_{pulse}/P_{pulse} =
1.2 \times 10^{-4}$ yr$^{-1}$ over the 600-d time interval.  The
simulated time series of accretion torque noise was then integrated
twice in succession in order to obtain its effect on pulse phase as a
function of time.

Each simulated data set was analyzed in the same way as the actual
\RXTE\ data, i.e., we tested a range of trial orbital periods and, for
each one, carried out a five parameter fit which included constant,
linear, and quadratic terms, as well as orbital amplitude and phase.
The best fitting ``orbital solution'' was selected from among the
various trial orbital periods, subject to the constraint that the mass
function exceed a minimum value, i.e., $f(M) > 0.14$ M$_{\sun}$.  This
constraint eliminates solutions with implausible values of the mass
function.  In particular, if we assume that the stellar masses are 12
M$_{\sun}$ and 1.4 M$_{\sun}$, then $f(M) > 0.14$ M$_{\sun}$ for 97\% of
randomly chosen orientations of the orbital plane.  If the mass of the
optical star is greater than 12 M$_{\sun}$, then it is even less
likely that the mass function would violate this constraint.

The rms deviation of the simulated measurements from the best fit, the
corresponding mass function, and the corresponding orbital period were
obtained for each of $10^4$ simulations.  For about 30\% of these,
there was no solution for any orbital period in the range 30 to 600
days with $f(M) > 0.14$ M$_{\sun}$.  Most of the remaining
$\sim$\,70\% have best-fit orbital periods near 600 days, i.e.,
commensurate with the length of the simulated data train.  The results
for the $\sim$\,45\% of the simulations that satisfy the mass function
constraint and for which the best-fit orbital period $P_orb < 600$ d
are summarized in Figure 8, where projections of three combinations of
the parameters are shown.  If we restrict the ``acceptable fits'' to
those with orbital periods $<350$ days, a value both comfortably
longer than 250 days and shorter than the length of the simulated data
train, then only 6.2\% of the original simulations fall into this
category.  If we further require that the fits yield rms deviations
less than 30 s, a value just above that for the best circular-orbit
fit of the actual PCA timing observations, then only 0.3\% of the
simulated data sets produce acceptable results.  Finally, if we
require that the solutions also have positive values of
$\dot{P}_{pulse}/ P_{pulse}$, then only 0.2\% of the simulations are
completely ``acceptable''.  Thus, we conclude that there is only a
small chance that a random walk in pulse phase would yield
sinusoidally varying pulse arrival time delays similar to those we
have observed, and thereby mimic the Doppler delays from a plausible
orbit of the neutron star.

\subsection{Pulse Timing Analysis of \RXTE\ ASM Data}

The intensities measured with the ASM were Fourier transformed after
the observation times were adjusted to the Solar System barycenter.
The resultant power density spectrum (PDS) is shown in Figure 9a for
periods in the vicinity of the 837-s pulse period; the average power
has been normalized to unity.  Note that there is a statistically
significant group of small peaks with powers exceeding $\sim$\,8 and
covering the period range of 837.3 to 838.0 s.  The observation times
were then further adjusted to remove the effects of an assumed
constant spin-down of the X-ray pulsar with a rate of
$\dot{P}_{pulse}/P_{pulse} = 1.56 \times 10^{-4}$ yr$^{-1}$ (see
Table 3).  The PDS of the data set with this correction is shown in
Figure 9b.  The power is now more concentrated in a narrow region
centered around a period of 837.3 s, with several peaks exceeding 15.
Finally, we corrected the observation times to remove the time delays
due to the eccentric orbit derived from the PCA analysis (see Table
2).  The results in Figure 9c now show a sharp peak (of height 76) at
837.33 sec, which is just the expected pulse period for the start of
the ASM epoch $\sim$\,2.5 yr earlier than the start of the PCA
observations.  \placetable{table3}

Given this clear detection of the pulsations in the ASM data, we
decided to utilize these data to independently determine the orbital
parameters.  To accomplish this, each of the parameters $P_{orb}$,
$a_{x} \sin i$, $T_{\pi/2}$, $e$, and $\dot{P}_{pulse}$, was varied
systematically over a wide range while the others were fixed at values
determined either in the PCA analysis or earlier in this procedure.
For each set of the parameter values, the ASM observation times
(corrected to the Solar System barycenter) were adjusted to remove the
effects of pulsar spin down and orbital motion so as to attempt to
obtain coherence in the 837-s pulsations.  Each adjusted set of data
was subjected to a Fourier analysis.  We then searched all the PDSs
made while varying a particular parameter to find the largest value of
the power, and to thereby establish a value for that parameter.  Each
of the parameters was varied in turn until the best values for all
were obtained.  Note that the pulse frequency was effectively varied
since we searched an extended frequency range ($858^{-1}$ - $770^{-1}$
Hz) of each PDS.  Rough estimates of the uncertainties in the
parameter values were established by finding the ranges of parameter
values that yielded maximum powers of at least $[P_{max}^{1/2} -
(1/2)^{1/2}]^2$, where $P_{max}$ is the largest power in any of the
transforms.

The orbital parameter determinations from the ASM data were carried
out for each of three separate sets: (i) data from MJD 50087 to 50995
(i.e., prior to the PCA observations), (ii) data from MJD 50995 to 51535
(i.e., contemporaneous with the PCA observation interval), and (iii) all
the ASM data.  The best fit parameters and their $1 \sigma$
single-parameter confidence errors are listed in Table 3.  Three
conclusions can be drawn from these numbers.  The first is that the orbital
parameter values are clearly consistent with those found from the PCA
analysis, though the uncertainties are larger for the ASM results.
Second, the value of $\dot{P}_{pulse}/P_{pulse}$ during the ASM
observations which preceded the PCA observations, was significantly larger
(by $\sim$\,20\%) than during the PCA observation interval.  Finally, the
existence of the 250-day orbit in X Per is clearly confirmed by the earlier
ASM observations which are completely independent of the PCA observations
(i.e., different detectors and non-overlapping time intervals).

\subsection{Search for Orbital Light Curve in the X-Ray Intensity Data}

If the neutron star moved through a stellar wind whose intensity falls
off as $1/r^2$, then a periodic variation in accretion rate that
ranges from $\dot{M}_0 {(1-e)}^{-2}$ to $\dot{M}_0 {(1+e)}^{-2}$ would
be expected from an orbit with a small eccentricity $e$, where
$\dot{M}_0$ is the mean accretion rate.  We therefore carried out a
formal search for sinusoidal variations in intensity as a function of
trial orbital period over the range 50 days to 600 days using the data
shown in Figure 2. No statistically significant periodicities were
found, and $2 \sigma$ upper limits on the amplitudes of such
periodicities were set for each trial period; the results are shown in
Figure 10. The $2 \sigma$ upper limit on periodic variability (as the
amplitude of a sine wave) in the vicinity of $P_{orb} \sim\,250$ d is
only 9\% of the mean count rate.  When this upper limit is interpreted
in terms of allowed orbital eccentricities via the above expressions,
the $2 \sigma$ upper limit on $e$ from the X-ray intensity data alone,
for orbital periods near 250 days is $e < 0.04$. This is clearly
inconsistent with the value obtained from the orbital fit of $e =
0.111 \pm 0.018$.  If, on the other hand, the stellar wind density
profile from the Be star falls off more slowly than $1/r^2$, the limit
that can be obtained from the intensity data alone is weaker, e.g.,
the limit would be $e < 0.09$ for a $1/r$ density profile.  Such a
value would be marginally consistent with the observed eccentricity.
Of course, the interpretation of the lack of X-ray variability, in
terms of orbital eccentricity, is based on the Bondi-Hoyle theory of
accretion from a stellar wind (\cite{wan81,liv96,ruf99}) which must be
considered substantially uncertain.  For a more extensive discussion
of neutron stars in eccentric orbits accreting from winds of Be star
companions, see Waters et al. (1988, 1989).

\section{DISCUSSION}

Bildsten et al. (1997) list 5 Be/X-ray binaries with measured orbital
parameters.  These are shown schematically, but to scale, in Figure
11.  For each system, the Be star is fixed, and the orbit of the
neutron star is drawn with the {\it full} semimajor axis $a$.  To
estimate the semimajor axis we utilized the measured {\em projected}
semimajor axis and mass function for the neutron star, and then simply
assumed masses for the Be and neutron stars of 18 M$_{\sun}$ and 1.4
M$_{\sun}$, respectively.  Also shown on the figure for comparison is
the X Per system.

The average eccentricity for the 5 Be/X-ray binaries other than X Per
is $e = 0.4$, in sharp contrast with the much smaller value of 0.11
for X Per.  This is especially noteworthy in that the X Per system has
the smallest value of $R_{Be}/a\ (\lesssim 0.05)$, which indicates that there
is virtually no chance that the system has circularized significantly
since the birth of the neutron star (see, e.g.,
\cite{zah77,zah89,ver95,sok98}).  Here we take 20 R$_{\sun}$ as the
upper limit on the radius of the Be star.  Thus, for any plausible
circularization time for such a Be star filling its Roche lobe (e.g.,
a few thousand yr), the tidal circularization timescale, which is
proportional to $(a/R)^8$, in the widely separated X Per system would
exceed the lifetime of X Per by orders of magnitude.

If we assume that the orbit of X Per has an eccentricity of 0.11 which
has remained unchanged since the birth of the neutron star, then we
can set interesting and potentially important constraints on the
dynamics of the supernova explosion and the mass that was ejected from
the binary system.  First, however, we consider the origin of the X
Per/4U0352+30 system.  In the conventional scenario for forming a Be
X-ray binary (\cite{rap82,vdh87,hab87}), the more massive star (e.g.,
$\sim$\,10-25 $\rm{M}_{\sun}$) in the progenitor binary evolves first
and fills its Roche lobe.  Depending on the evolutionary state of the
primary at the onset of mass transfer (see, e.g., \cite{kip67,pod92}
and references therein), the Roche lobe overflow of mass onto the
secondary may proceed either on a thermal timescale or be dynamically
unstable.  In the former case, relevant to the formation of Be stars,
it is unclear how conservative the mass transfer may be.  In general,
the orbital separation will first decrease, and then, if conditions
are right (see below), increase until the mass transfer terminates.

In this scenario, detailed binary evolution calculations are required
to determine how much envelope mass is retained by the He core of the
original primary star (see, e.g., Podsiadlowski et al. 1992).
However, simple orbital dynamics provide an analytic relation among
the final orbital separation, the initial separation, and the final
and initial masses of the system (see equation 5 of Podsiadlowski et
al. 1992).  There are two ``free'' parameters in this relation, viz.,
$\alpha$, the specific angular momentum carried away by matter ejected
from the binary system in units of the binary angular momentum per
reduced mass, and $\beta$, the fraction of mass lost by the primary
that is retained by the secondary during mass transfer.  We plot in
Figure 12 some illustrative examples of the ratio of final to initial
orbital separations as a function of fractional mass loss of the
primary star for different assumed values of $\beta$.  The initial
stellar masses in this particular example are 15 and 10
$\rm{M}_{\sun}$, and we adopted a value $\alpha = 1.5$ which might be
typical for mass lost through the outer Lagrange points.  (We remind
the reader that in the above paragraph as well as in the next
paragraph the words ``initial'' and ``final'' refer only to the
process of quasi-conservative mass transfer from the primary to the
secondary {\em prior} to the supernova.)

As one can see from Figure 12, for conservative mass transfer the
final orbit can be substantially larger than the initial orbit.
However, for $\beta \lesssim 0.6$ the orbit tends to shrink
dramatically.  This is true for a range of plausible initial mass
ratios.  If the initial orbital separation in the X Per system had
been larger than $\sim\,5$ AU, the primary would not have filled its
Roche lobe and transferred mass to the secondary (and the system would
have become unbound after the supernova). Moreover, the orbit just
prior to the supernova explosion was probably not much closer than 2
AU (see Fig. 14 and the associated discussion below).  Thus, since the
orbit did not shrink by more than a factor of $\sim\,5/2$ during the
mass-transfer phase from the primary, we can see from Fig. 12 that
$\beta$ cannot be substantially less than $\sim\,0.6$.  Finally, we
conclude that the initial orbital separation of the primordial binary
probably could not have been much smaller than about 0.5 AU. This
latter conclusion is based upon the fact that in order for the final
to initial orbital separation to increase by a factor of $\sim$\,4,
the value of $\beta$ must be close to unity; then one must also
satisfy the condition that the computed current-epoch mass of X Per
should not be too large.  In summary, we infer that the primordial
binary which formed the current X Per system had an initial orbital
separation of between 0.5 and 5 AU, experienced quasi-stable
Roche-lobe overflow when the primary evolved, and that most of the
mass lost by the primary was retained by the secondary.

We now consider the effects on the orbit of the collapse of the Fe
core of the remnant primary and the ensuing supernova explosion.
First, we examine the case where the supernova explosion was
spherically symmetric in the rest frame of the progenitor star, i.e.,
no kick velocity was imparted to the newly born neutron star.  If the
orbit of the progenitor of the neutron star and its companion (i.e.,
the present-day X Per) was circular, and there was no kick imparted to
the neutron star at its birth, then there is a simple relationship
among the current orbital eccentricity, the total mass of the
progenitor system, $M_{tot}$ (just before the supernova explosion),
and the mass ejected by the explosion, $\Delta M$:
\begin{equation}
          e = \Delta M/(M_{tot} - \Delta M)
\end{equation}
Some illustrative examples taken from this relationship are plotted in
Figure 13.  Here we show the eccentricity induced by the supernova
explosion as a function of the mass ejected, for a plausible range of
values (10 - 30 M$_{\sun}$) for the total mass of the binary just
prior to the supernova explosion.  Note that for the supernova to have
produced an orbital eccentricity $e < 0.15$ (i.e., the $2\sigma$
observational upper limit), no more than $\sim4\,\rm{M}_{\sun}$ can
have been ejected from the system.  Thus, we conclude that the
progenitor of the neutron star was a He star with mass $<
5-6\,\rm{M}_{\sun}$. Estimates of the minimum mass of a He star that
can produce a neutron star are close to 2.3\,M$_{\sun}$ (see, e.g.,
\cite{hab86a,hab86b}).

The results shown in Figure 13 are for the case where the neutron star
was {\em not} given a kick when the Fe core of its progenitor star
collapsed.  However, it is conventional wisdom that most neutron stars
are given substantial kicks at their birth (e.g.,
\cite{ll94,cor97,cor98,fry98,han97}).  The measured 3-dimensional
space velocity distribution for isolated neutron stars is uncertain,
but is generally characterized by a function with a mean speed of
$\sim$\,300 km $s^{-1}$ and which tends to vanish at low speeds.
There seems to be a growing consensus that these large space
velocities result from natal kicks to the neutron star, and that the
``slingshot'' effect from neutron stars born in binary systems is
insufficient to produce the observed velocities (see, e.g., Cordes \&
Chernoff 1997, 1998; Fryer, Burrows, \& Benz 1998; Hansen \& Phinney
1997; cf. \cite{ibe96}).  The small observed eccentricity in X Per
seems to be qualitatively at odds with this ``universal'' kick
distribution.

In order to study the problem more quantitatively, we have carried out
a modest Monte Carlo study of supernovae in binary systems of the type
that are directly relevant to the formation of 4U~0352+30/X Per.  We
considered circular orbits for the progenitor He star and its massive
companion (the current X Per).  We chose the mass of the He star to be
6 M$_{\sun}$. For more massive He stars even a no-kick supernova
explosion would leave the orbit more eccentric than is observed --
unless the supernova kick fortuitously ``corrects'' the eccentricity
that would otherwise be induced.  We take the mass of the companion
star to be 18 M$_{\sun}$, which is approximately the current mass of X
Per.  We also considered 4 different initial orbital separations for
the pre-supernova binary: 0.5, 1, 2, and 4 AU.  For each Monte Carlo
case we chose at random a kick speed from a Maxwellian distribution
(Hansen \& Phinney 1997):
\begin{equation}
    p(v) = \sqrt{ \frac{\pi}{2} } \frac{v^2}{v_0^3} e^{-v^2/2v_0^2},
\end{equation}
while the direction of the kick was chosen from an isotropic
distribution (see, e.g., \cite{bra95}). In this expression we adopted
a value for $v_0$ of 190 km s$^{-1}$ which yields a mean kick velocity of
320 km s$^{-1}$ (\cite{han97}).  These quantities then uniquely
determine the final post-supernova orbital parameters; these were
recorded for each system.  For each separation this procedure was
repeated $10^6$ times, and the distributions of the results are shown
in Figure 14.  We find that $\sim$\,3-24\% of the binaries remain
bound following the supernova explosion, with the exact percentage
depending sensitively on the pre-supernova semimajor axis.  The best
match to the current orbital period of 250 days is an initial orbital
separation of 2 AU.  However, as can readily be seen from the
distribution of eccentricities, there is only a {\it very small}
probability of finding the post-supernova orbit with an eccentricity
of $\lesssim$\,0.11; typically $\sim$\,1\% of the binaries which remain
bound have an eccentricity this small.

From our analysis it seems reasonable to conclude that, at least for
the X Per binary, the Be/neutron star pair formed via a completely
unremarkable scenario.  Because of the small eccentricity in the
system, however, our Monte Carlo kick study clearly indicates that it
is unlikely that a substantial kick was imparted to the neutron star
at birth.  However, one should be cautious about drawing far reaching
conclusions based on this one system.  We point out that there are at
least two other relatively wide X-ray binaries that may exhibit a
similarly small orbital eccentricity: 2S 1553--54 and GS 0834--43
(see \cite{bil97} and references therein).  The orbit of the transient
X-ray source 2S 1553--54 was determined on only one occasion during a
single orbital cycle (\cite{kra83}).  The system was found to have a
period of 29 days and an eccentricity $e < 0.09$ (2$\sigma$ limit).
The companion star has never been optically identified.  Nonetheless
this system seems like another reasonable candidate for a Be/X-ray
binary with only a small eccentricity.  The transient source GS
0834--43 (\cite{wil97}) was found to have an orbital period of 106
days, but the orbital parameters were not uniquely determined due to
effects of accretion torque noise.  However, the best fitting
solutions yield an orbital eccentricity of $\sim$\,0.17.  The optical
counterpart of GS~0834--43 has recently been identified with a Be
star (\cite{isr00}).

While it is always conceivable that this system was produced via a
highly improbable event, it seems more reasonable in the case of X Per
to draw the conclusion that not all neutron stars receive kicks of a
magnitude consistent with the Hansen \& Phinney (1997) distribution of
single pulsar velocities.  Therefore we adopt the somewhat
unconventional view that perhaps a substantial fraction of all neutron
stars do {\em not} suffer significant kicks at birth (see also
\cite{ibe96,cor98}).

\acknowledgements

This work was supported in part by NASA Grants NAG5-7479, NAG5-4057,
and NAG5-8368, and Contract NAS5-30612.  We are grateful to Jean Swank
and Evan Smith for support in conducting the \RXTE\ observations.

\begin{deluxetable}{cccccc}
\tablewidth{0pt}
\tablecaption{Journal of Observations\label{table1}}
\tablehead{
\colhead{Start Time} &
\colhead{Duration} &
\colhead{Exposure} &
\colhead{Count Rate\tablenotemark{a}} &
\colhead{Pulse Arrival Time\tablenotemark{b}} &
\colhead{Pulse Number}  \\
\colhead{(MJD\tablenotemark{c}\ )} & \colhead{(s)} &
\colhead{(s)} &
\colhead{(cts s$^{-1}$ PCU$^{-1}$)} &
\colhead{(s)} & \colhead{}
}
\startdata
50995.033 &     7648 &  4960 &  34.4 & 0 &          0 \nl
50996.651 &     7472 &  5088 &  32.1 & 139033 &     166 \nl
50998.846 &     7631 &  5056 &  25.1 & 329189 &     393 \nl
51000.897 &     3327 &  3328 &  24.1 & 505881 &     604 \nl
51002.963 &     6944 &  4464 &  34.5 & 684357 &     817 \nl
51005.925 &     11056 & 6256 &  28.7 & 940623 &     1123 \nl
51008.031 &     8255 &  5648 &  28.7 & 1122409 &    1340 \nl
51011.030 &     7231 &  4912 &  25.3 & 1382033 &    1650 \nl
51014.775 &     8191 &  5360 &  31.8 & 1705366 &    2036 \nl
51018.774 &     7232 &  4544 &  27.8 & 2050419 &    2448 \nl
51021.829 &     7712 &  5520 &  26.0 & 2315078 &    2764 \nl
51025.831 &     7344 &  5296 &  24.8 & 2660167 &    3176 \nl
51029.998 &     6864 &  4448 &  27.4 & 3020347 &    3606 \nl
51034.769 &     6928 &  4944 &  33.6 & 3432402 &    4098 \nl
51039.799 &     7023 &  4864 &  35.6 & 3867102 &    4617 \nl
51044.805 &     10176 & 5328 &  26.5 & 4300151 &    5134 \nl
51049.706 &     7664 &  5264 &  28.8 & 4723142 &    5639 \nl
51054.652 &     7696 &  4864 &  24.2 & 5150307 &    6149 \nl
51060.839 &     7295 &  4907 &  47.0 & 5685524 &    6788 \nl
51066.506 &     8352 &  4832 &  35.1 & 6174684 &    7372 \nl
51072.704 &     7296 &  5136 &  14.4 & 6710771 &    8012 \nl
51079.372 &     7936 &  3936 &  22.4 & 7286185 &    8699 \nl
51085.645 &     7424 &  5488 &  31.9 & 7828195 &    9346 \nl
51092.540 &     6847 &  4592 &  34.5 & 8423815 &    10057 \nl
51099.711 &     8127 &  5440 &  33.7 & 9043673 &    10797 \nl
51107.576 &     7583 &  4832 &  23.4 & 9723109 &    11608 \nl
51115.594 &     7567 &  5152 &  34.8 & 10415978 &   12435 \nl
51122.445 &     8415 &  5488 &  31.0 & 11008235 &   13142 \nl
51131.502 &     7231 &  4880 &  33.4 & 11790789 &   14076 \nl
51139.718 &     8431 &  5184 &  37.4 & 12500405 &   14923 \nl
51147.714 &     8542 &  5408 &  33.4 & 13191633 &   15748 \nl
51154.436 &     7583 &  5200 &  23.5 & 13772213 &   16441 \nl
51166.297 &     8654 &  5343 &  24.9 & 14796872 &   17664 \nl
51174.626 &     7167 &  4880 &  30.4 & 15516615 &   18523 \nl
51185.287 &     7711 &  4896 &  36.0 & 16437345 &   19622 \nl
51194.221 &     7663 &  4960 &  25.8 & 17209750 &   20544 \nl
51204.417 &     8078 &  5375 &  33.1 & 18090212 &   21595 \nl
51214.344 &     7231 &  4912 &  29.2 & 18948065 &   22619 \nl
51225.268 &     8014 &  5664 &  26.9 & 19892186 &   23746 \nl
51236.063 &     8414 &  4768 &  22.8 & 20824568 &   24859 \nl
51247.267 &     6510 &  4480 &  28.1 & 21792933 &   26015 \nl
51258.118 &     7182 &  4896 &  34.4 & 22730264 &   27134 \nl
51269.310 &     7518 &  4880 &  34.2 & 23696849 &   28288 \nl
51281.164 &     8014 &  5808 &  24.0 & 24721370 &   29511 \nl
51292.888 &     6767 &  3360 &  43.4 & 25734120 &   30720 \nl
51355.004 &     6847 &  4640 &  20.9 & 31101333 &   37127 \nl
51367.899 &     7344 &  4896 &  26.4 & 32214823 &   38456\nl
51377.888 &     7279 &  4976 &  27.3 & 33077855 &   39486 \nl
51392.937 &     3727 &  3712 &  32.0 & 34378244 &   41038 \nl
51406.007 &     8608 &  5328 &  27.1 & 35507717 &   42386 \nl
51420.662 &     7776 &  5288 &  44.7 & 36773788 &   43897 \nl
51433.721 &     9760 &  5154 &  34.5 & 37902423 &   45244 \nl
51447.614 &     8739 &  6258 &  22.1 & 39102200 &   46676 \nl
51463.877 &     8448 &  5244 &  23.7 & 40508074 &   48354 \nl
51478.579 &     7543 &  5130 &  22.6 & 41778194 &   49870 \nl
51492.622 &     8139 &  5958 &  24.8 & 42991263 &   51318 \nl
51512.878 &     19414 & 5165 &  25.1 & 44741312 &   53407 \nl
51534.688 &     8378 &  5778 &  35.7 & 46626158 &   55657 \nl
51556.300 &     8439 &  5838 &  24.2 & 48493430 &   57886 \nl
51576.474 &     7838 &  4939 &  27.8 & 50236005 &   59966 \nl
51597.240 &     7726 &  5306 &  38.4 & 52030604 &   62108 \nl
\enddata
\tablenotetext{a}{Average count rate for the energy band 2-20 keV}
\tablenotetext{b}{At the Solar System barycenter referenced to 1998 July
 1 0:33:6 TDB}
\tablenotetext{c}{Modified Julian Date $=$ Julian Date $- \ 2,400,000.5$}
\end{deluxetable}

\begin{deluxetable}{lcc}
\tablewidth{0pt}
\tablecaption{Orbital and Pulse Parameters\label{table2}}
\tablehead{
\colhead{Parameter} &
\colhead{Circular Orbit Fit} &
\colhead{Eccentric Orbit Fit}
}
\startdata
$P_{orb}$ (days) & $249.9 \pm 0.5$ & $250.3 \pm 0.6$ \nl
$a_{x} \sin i$ (lt-s) & $454 \pm 5$ & $454 \pm 4$ \nl
$f(M)$ (M$_{\sun}$) & $1.61 \pm 0.06$ & $1.61 \pm 0.05$ \nl
$T_{\pi/2}$ (MJD) & $51215.5 \pm 0.5$ & $51215.1 \pm 0.4$ \nl
$P_{pulse}$ (s)\tablenotemark{a} & $837.6713 \pm 0.0003$ & $837.6712 \pm 0.0003$ \nl
$\dot{P}_{pulse}/P_{pulse}$ (yr$^{-1}$) & $(1.26 \pm 0.01) \times 10^{-4}$ &
 $(1.26 \pm 0.01) \times 10^{-4}$ \nl
$e$           & \nodata & $0.111 \pm 0.018$ \nl
$\omega_{peri}$ ($\arcdeg$) & \nodata & $288 \pm 9$ \nl
$T_{peri}$ (MJD) & \nodata & $51353 \pm 7$ \nl
RMS Scatter (s) & 27.7 & 21.8 \nl
$a_0$\tablenotemark{b} & \nodata & 31055.6017 \nl
$a_1$\tablenotemark{b} & \nodata & $1.1936637 \times 10^{-3}$ \nl
$a_2$\tablenotemark{b} & \nodata & $-2.388 \times 10^{-15}$ \nl
\enddata
\tablenotetext{a}{The epoch for $P_{pulse}$ is MJD 51000.0.}
\tablenotetext{b}{The pulse phase at the Solar System barycenter may be obtained
from $\phi = a_0 + a_1 t + a_2 t^2 - \Delta t_{orb}(t)/P_{orb}$ where
$\phi$ is in cycles, $t$ is the time at the Solar System barycenter in
seconds since MJD 51296.12602, and $\Delta t_{orb}(t)$ is the light travel
time delay for the eccentric orbit relative to the X Per system
barycenter.  The phase is defined so that $\phi = 0$ at maximum intensity
(see Fig. 3).}
\end{deluxetable}

\begin{deluxetable}{lccc}
\tablewidth{0pt}
\tablecaption{Orbital Parameters Determined with ASM Data\label{table3}}
\tablehead{
\colhead{Parameter} &
\colhead{Pre-PCA Data\tablenotemark{a}} &
\colhead{Co-PCA Data\tablenotemark{b}} &
\colhead{All ASM Data\tablenotemark{c}}
}
\startdata
$P_{orb}$ (days) & $250 \pm 3$ & $253 \pm 10$ & $252 \pm 5$ \nl
$a_{x} \sin i$ (lt-s) & $421 \pm 70$ & $385 \pm 108$ & $444 \pm 66$ \nl
$T_{\pi/2}$ (MJD) & $51213 \pm 6$ & $51212 \pm 8$ & $51215 \pm 6$ \nl
$P_{pulse}$ (s)\tablenotemark{d} & $837.654 \pm 0.02$ & $837.658 \pm 0.01$ & $837.666 \pm 0.01$ \nl
$\dot{P}_{pulse}/P_{pulse}$ (yr$^{-1}$) & $(1.57 \pm 0.10) \times 10^{-4}$ &
 $(1.23 \pm 0.08) \times 10^{-4}$ & $(1.63 \pm 0.05) \times 10^{-4}$ \nl
$e$           & \nodata & \nodata & $< 0.3$ \nl
\enddata
\tablenotetext{a}{Data from time interval MJD 50087 to MJD 50995}
\tablenotetext{b}{Data from time interval MJD 50995 to MJD 51535}
\tablenotetext{c}{Data from time interval MJD 50087 to MJD 51535}
\tablenotetext{d}{The epoch for $P_{pulse}$ is MJD 51000.0.}
\end{deluxetable}

\begin{figure*}
\figurenum{1}
\plotfiddle{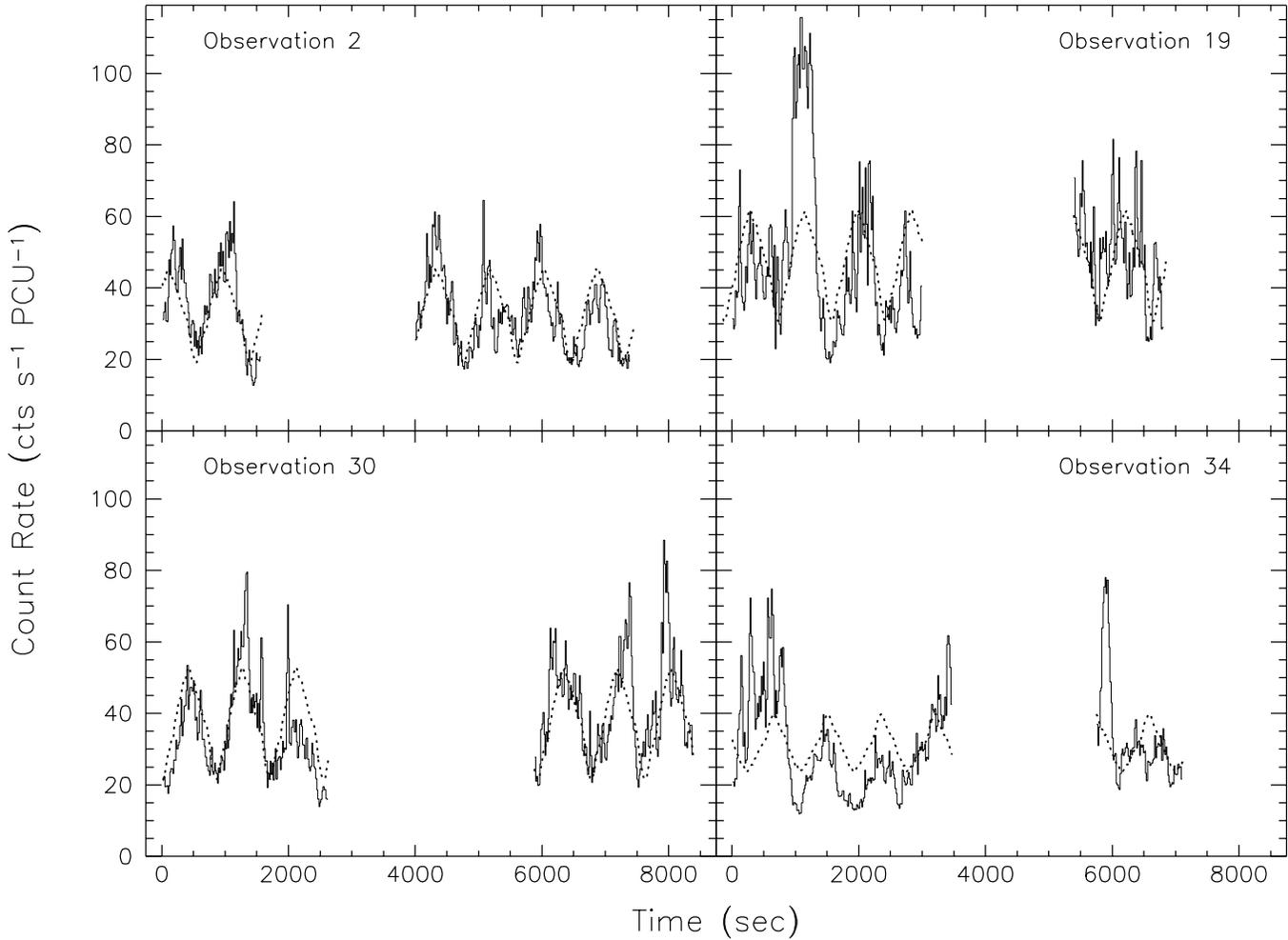}{6in}{-90.0}{70}{70}{-285}{450}
\caption{Counting rates (2 - 20 keV) during 4 of the
61 PCA observations of 4U 0352+30/X Per.  Non-source background has
been subtracted.  Each observation lasted for typically 7000 s, and
was interrupted by a $\sim$\,2000-s occultation of the source by the
Earth.  The dashed curves were derived from the average pulse profile
(see Figure 2), but a scale factor and additive constant were adjusted
for each observation to provide the best fit.  These scaled average
profiles are meant only to guide the eye and to indicate the degree of
variability in the pulsations. \label{fig1}}
\end{figure*}

\begin{figure*}
\figurenum{2}
\plotfiddle{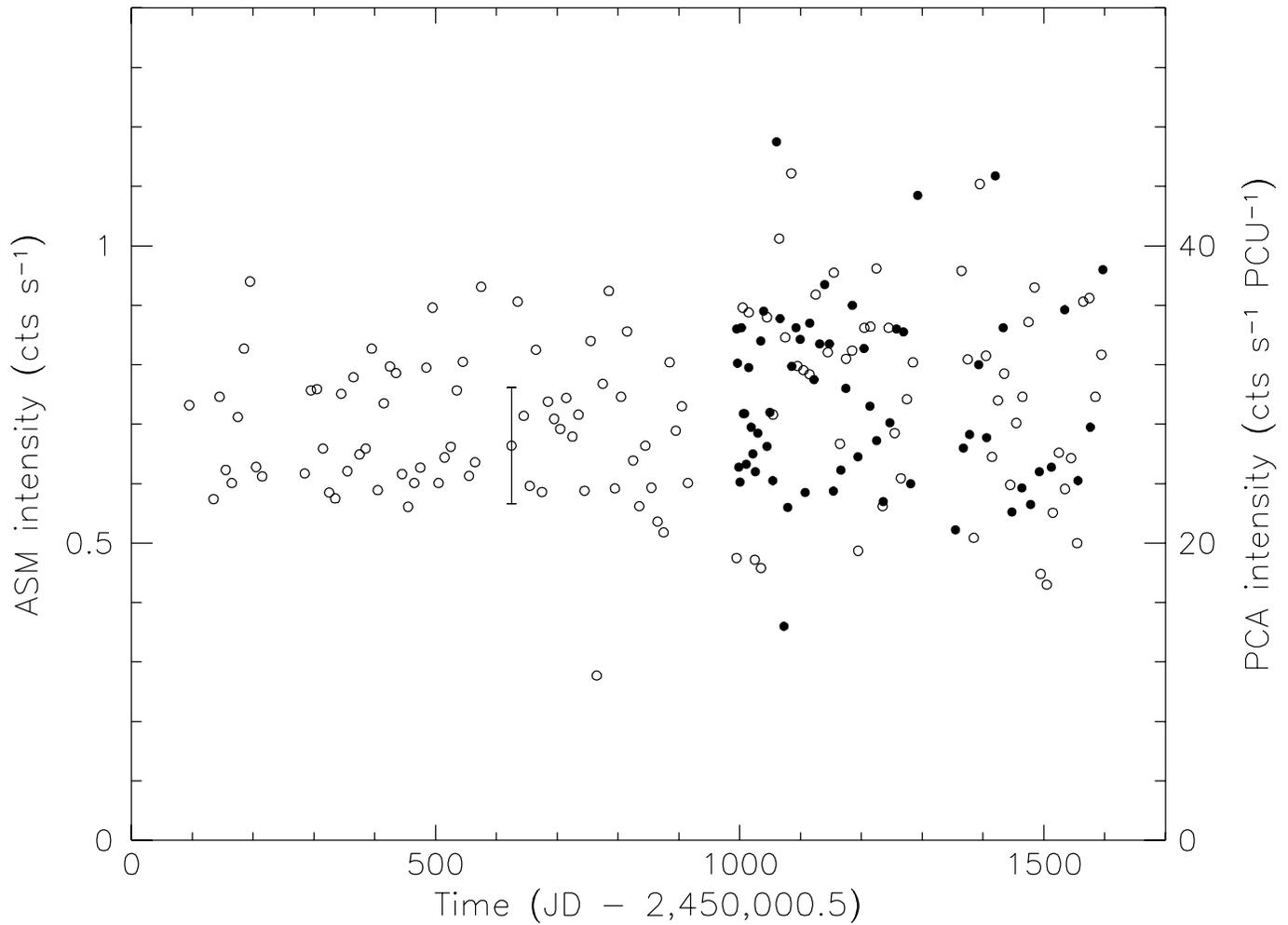}{6in}{-90.0}{70}{70}{-285}{450}
\caption{
X-ray intensity data for 4U 0352+30 over a 4-year interval.  The open
circles are 10-day averages (2--12 keV) from the ASM (scale on left),
while the filled circles are 2--20 keV average background-subtracted
count rates from each of the PCA observations (scale on right). A
typical $1 \sigma$ uncertainty is shown for one of the ASM
intensities.  Errors in the average PCA intensities due to counting
statistics are negligible.
\label{fig2}}
\end{figure*}

\begin{figure*}
\figurenum{3}
\epsscale{1.0}
\plotone{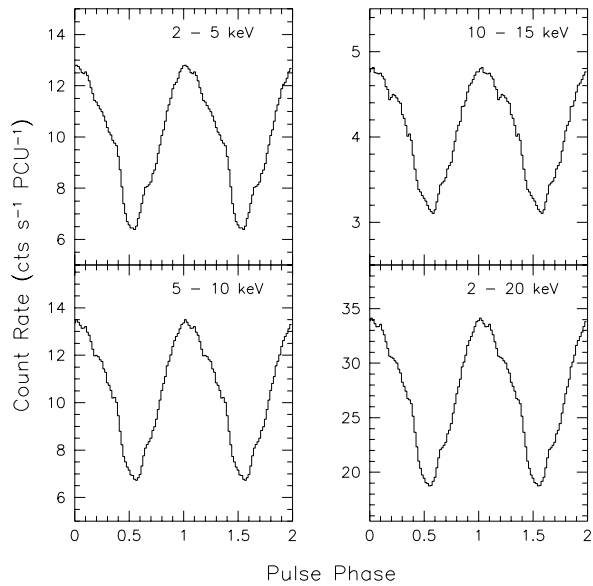}
\caption{ 
Background-subtracted pulse profiles for 4U~0352+30 in 4 energy bands.
The profiles are the averages of the phase-aligned pulse profiles from
each of the 61 PCA observations.  The fluctuations in the profiles due
to counting statistics are negligible.  Other fluctuations due to the
flaring behavior of the source on $\sim$\,100 s time scales may not be
negligible.
\label{fig3}}
\end{figure*}

\begin{figure*}
\figurenum{4}
\plotfiddle{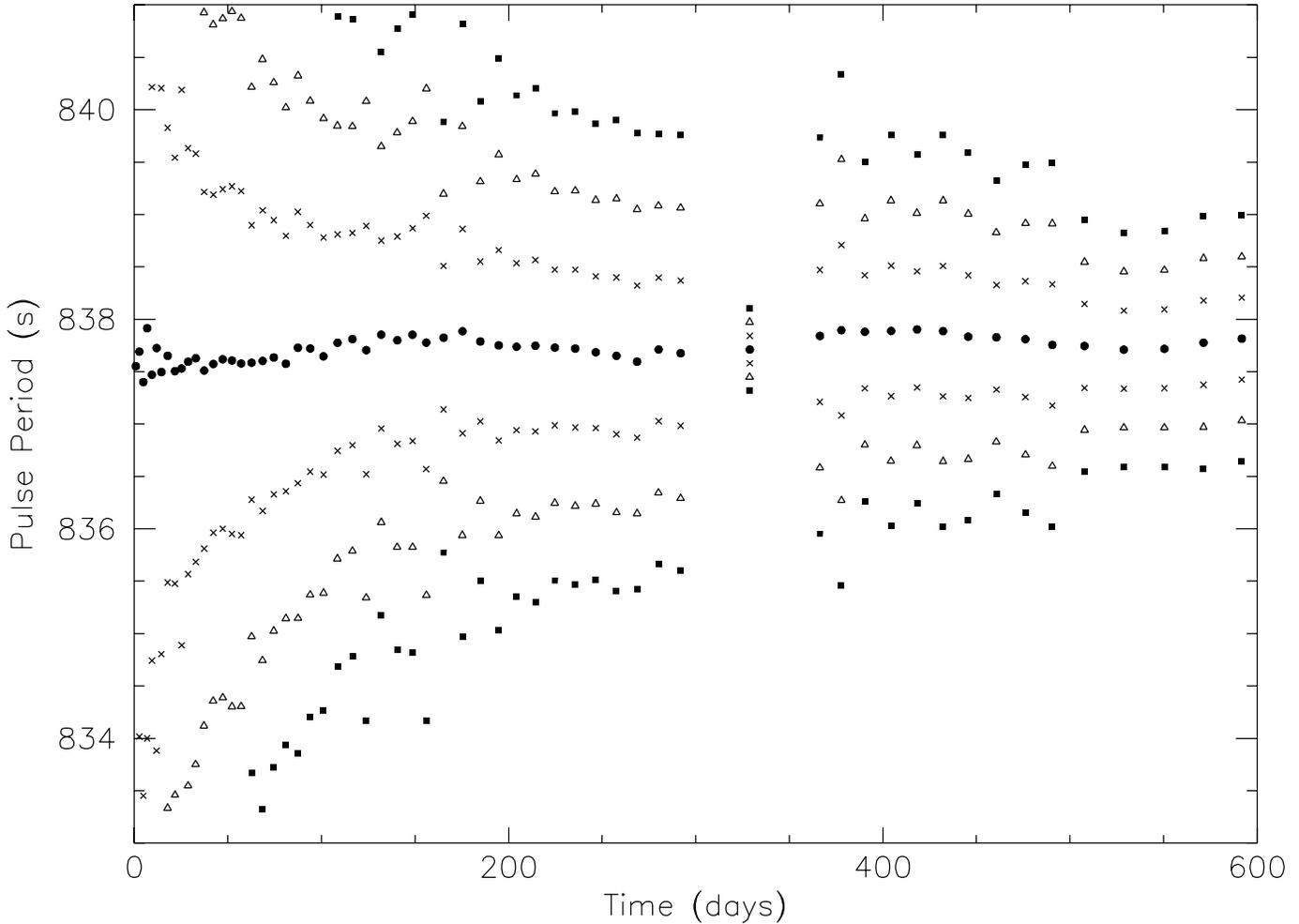}{5in}{-90.0}{70}{70}{-285}{400}
\caption{Average inter-observation pulse periods for 4U 0352+30.
Each pulse period was derived by taking the difference in pulse
arrival times between adjacent observations and dividing by an assumed
number of pulses between the observations.  Pulse periods are shown
for seven possible values of the number of pulses between each pair of
observations.  Only for the assumed pulse numbers that yield the
filled circles does the pulse period change slowly and remain within
the physically plausible range of $837.7 \pm 0.5$ s.  The only
possible ambiguity in pulse number occurs near the large gap in the
observations centered near day 330 (see text). \label{fig4}}
\end{figure*}

\begin{figure*}
\figurenum{5}
\epsscale{1.0}
\plotone{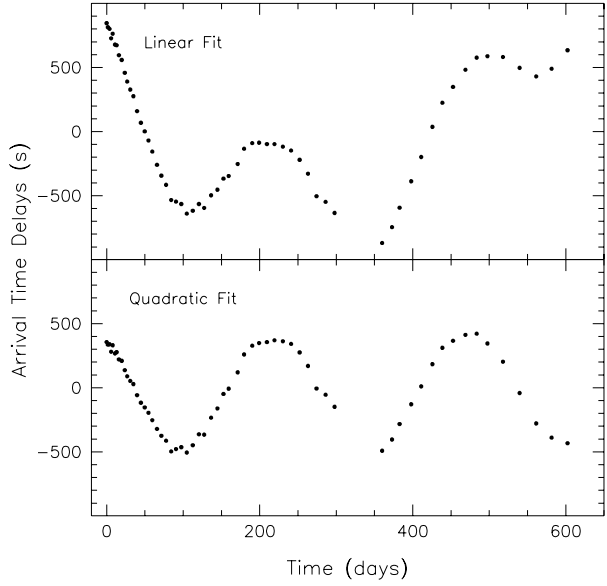}
\caption{ 
Pulse arrival time delays for 4U~0352+30 with respect to the best-fit
constant pulse period (top), and with respect to the best-fit
quadratic function (bottom).
\label{fig5}}
\end{figure*}

\begin{figure*}
\figurenum{6}
\epsscale{1.0}
\plotone{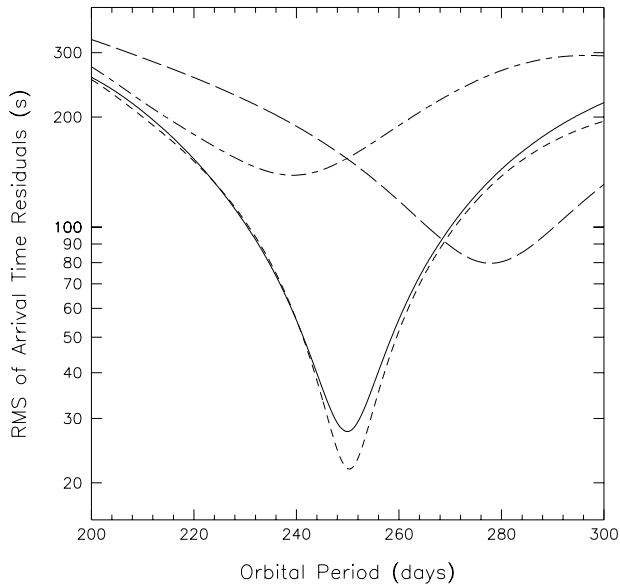}
\caption{ 
RMS residuals from fits of pulse arrival times vs. trial orbital
period.  The results are shown for the circular orbit fits (solid
line), and for mildly eccentric fits with the pulse count in the 62-d
gap having three different values, i.e, our nominal count $n$ (short
dashes), $n+1$ (long dashes), and $n-1$ (short/long dashes).  These
results confirm that $n$ is the correct pulse count for the gap.  The
best fitting orbital period is close to 250 d (see Table 2).
\label{fig6}}
\end{figure*}

\begin{figure*}
\figurenum{7}
\epsscale{1.0}
\plotone{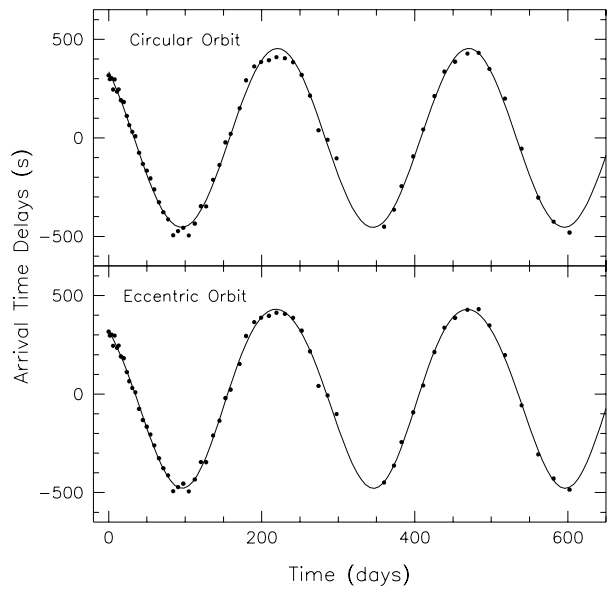}
\caption{ 
Pulse arrival time delays for 4U~0352+30 with fits to a quadratic
function plus a circular orbit (top) or a mildly eccentric orbit
(bottom).  In each case, the best fit quadratic function has been
subtracted from the pulse arrival times.  The solid curves are the
best fit model orbital Doppler delays.
\label{fig7}}
\end{figure*}

\begin{figure*}
\figurenum{8}
\plotfiddle{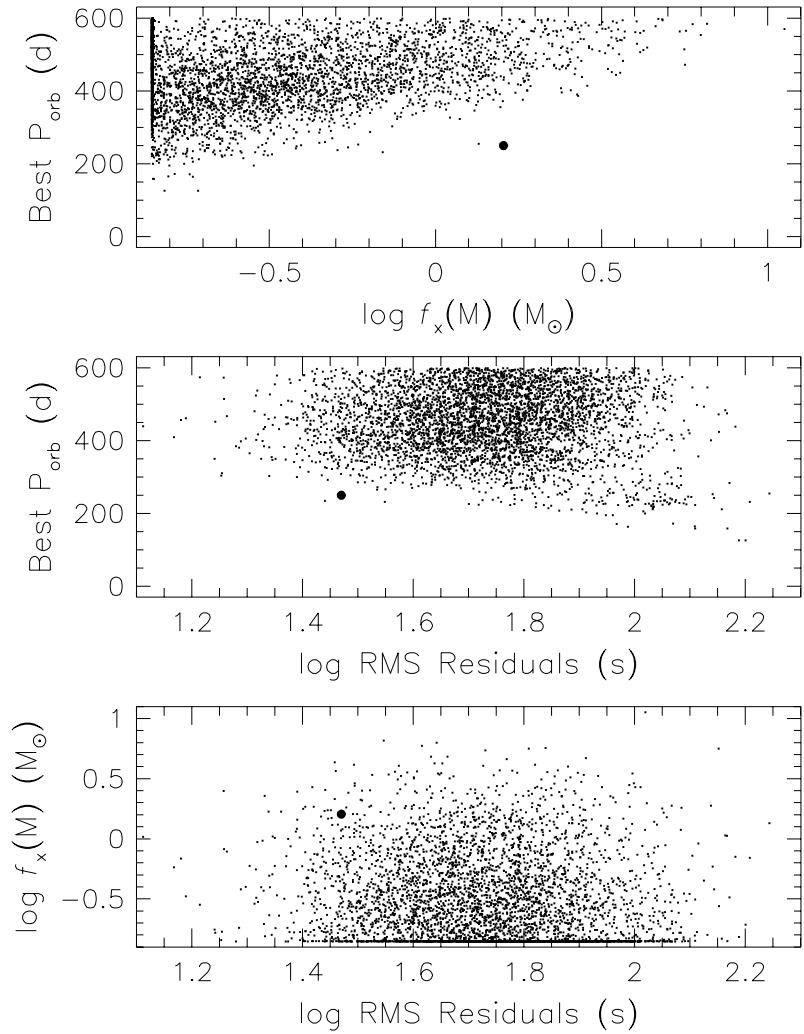}{6in}{0.0}{55}{55}{-150}{0}
\caption{
Results from $10^4$ simulations of a pulsar subject to white
torque noise. In each panel, each small dot shows the values of two
parameters from the best circular-orbit fit to one simulated data set
(see text).  We only show the dots from the 4490 simulations for which
the best fit orbital period was less than 600 days and for which $f(M)
> 0.14$ M$_{\sun}$.  The large filled circles show the values of the
parameters determined from the best circular orbit fit to the actual
measurements of 4U~0352+30 (see Table 2).
\label{fig8}}
\end{figure*}

\begin{figure*}
\figurenum{9}
\plotfiddle{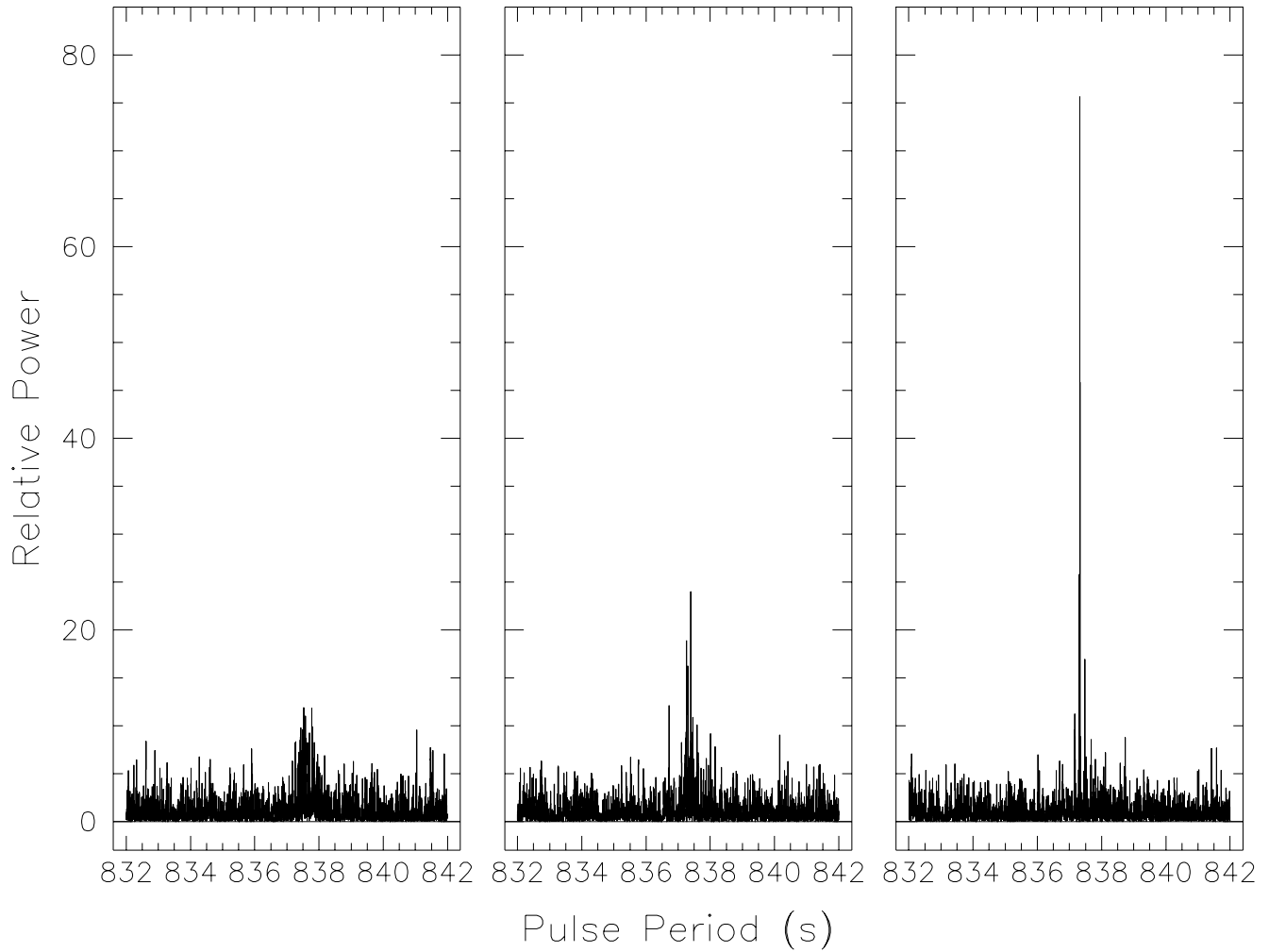}{6in}{-90.0}{70}{70}{-285}{450}
\caption{
Power density spectra (PDSs) computed from data obtained with the ASM.
Approximately 21,000 intensity points for 4U 0352+30, obtained over a
4 year interval, were used in the analysis.  (left) PDS which results
after barycentric corrections were applied to the data.  (center) PDS
after correcting the times to the barycenter and for a constant value
of $\dot{P}_{pulse}/P_{pulse} = 1.56 \times 10^{-4}$ yr$^{-1}$.
(right) PDS after corrections for the eccentric orbit given in Table 2
in addition to the barycentric and pulse period derivative
corrections.
\label{fig9}}
\end{figure*}

\begin{figure*}
\figurenum{10}
\epsscale{1.0}
\plotone{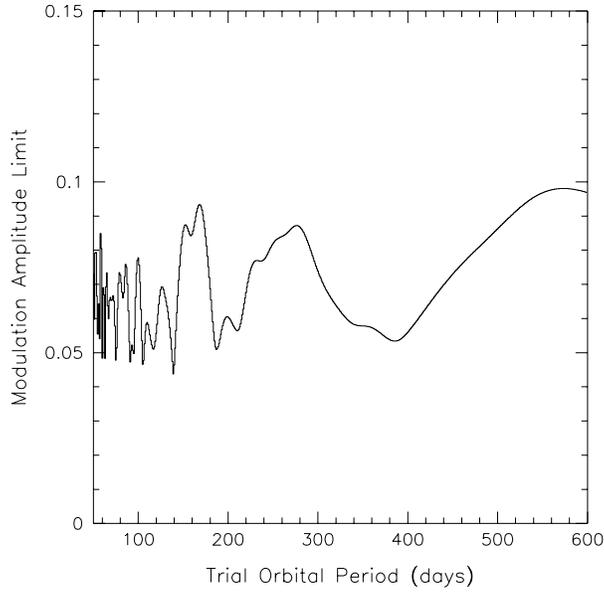}
\caption{
Upper limits on sinusoidal variations in the X-ray intensity of 4U
0352+30 derived from the data shown in Figure 2 as a function of trial
orbital period.  The limits are expressed as fractions of the average
intensity and are shown for the $2\sigma$ confidence level.
\label{fig10}}
\end{figure*}

\begin{figure*}
\figurenum{11}
\epsscale{1.0}
\plotone{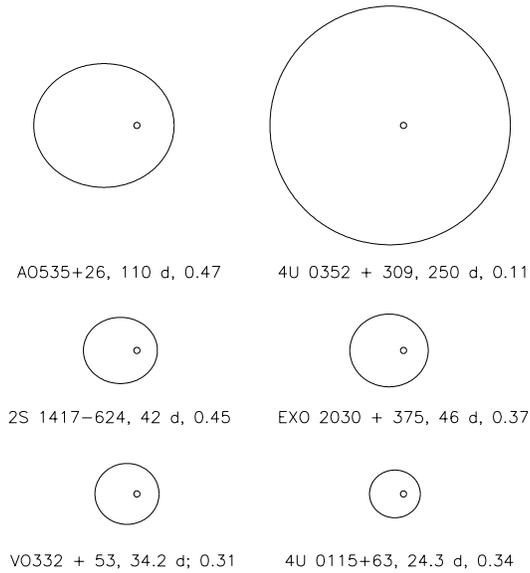}
\caption{
Schematic orbits for 6 Be/X-ray binaries drawn to scale (see Bildsten
et al. 1997). The orbital period and eccentricity are listed with the
source name.  The small circle represents the Be star with an
illustrative radius of 10 $R_{\sun}$.
\label{fig11}}
\end{figure*}

\begin{figure*}
\figurenum{12}
\epsscale{1.0}
\plotone{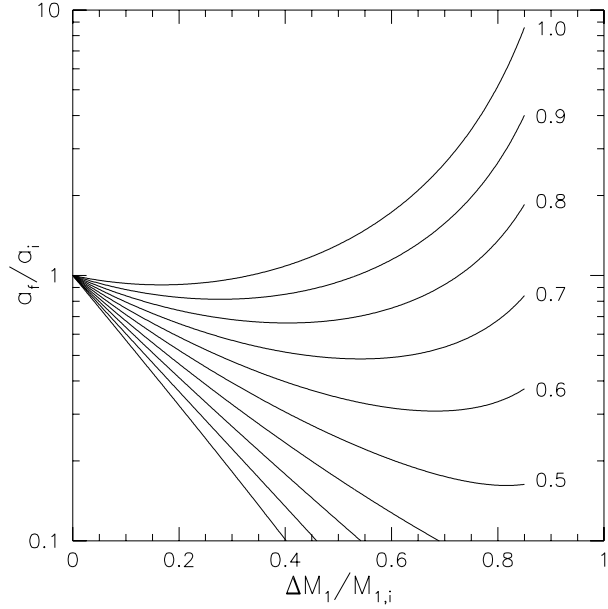}
\caption{
Ratio of final to initial pre-supernova orbital separation vs. the
fractional mass lost, $\Delta M_1/M_{1,i}$, by the primary for the
case where a fraction $\beta$ of mass lost by the primary is retained
by the secondary during the Roche-lobe overflow phase (see text).  The
curves are labelled according to the value of $\beta$ used in the
calculation.  The specific angular momentum of the mass ejected from
the binary system was taken to be $1.5 a^2 \Omega_k$, where $a$ and
$\Omega_k$ are the instantaneous values of the orbital separation and
Keplerian angular frequency, respectively.
\label{fig12}}
\end{figure*}

\begin{figure*}
\figurenum{13}
\epsscale{1.0}
\plotone{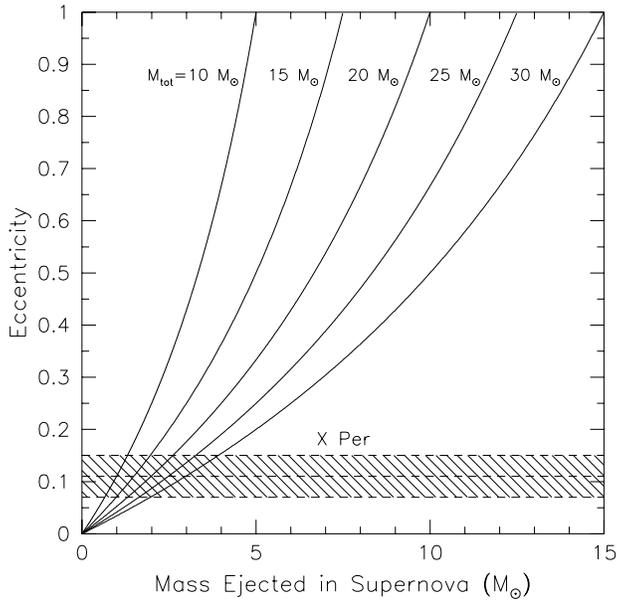}
\caption{
The eccentricity induced in binary systems as a function of the mass
ejected in a supernova explosion.  The calculations are for the case
where there is no natal kick imparted to the neutron star.  Each curve
is labeled with the total mass of the binary system prior to the
supernova explosion.
\label{fig13}}
\end{figure*}

\begin{figure*}
\figurenum{14}
\epsscale{1.0}
\plotone{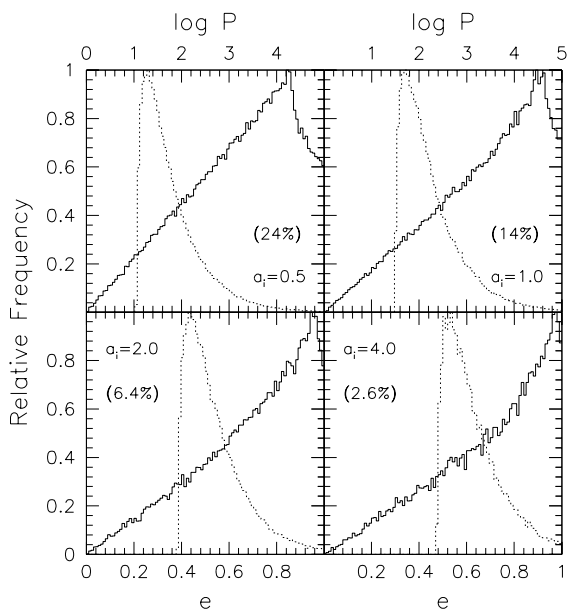}
\caption{ Results of a Monte Carlo study of the eccentricities induced
in binary systems in which kicks are imparted to the neutron star
during the supernova explosion.  Each panel shows the results from the
simulation of $10^6$ binary systems with a fixed initial (just before
the supernova) orbital separation $a_i$ (in AU).  A kick velocity
distribution given by equation (3) was used to select the kicks.  The
solid (dashed) curves are histograms of the final eccentricity
(logarithm of the orbital period in days).  The percentage of systems
that remain bound after the supernova explosion is also given for each
initial separation.
\label{fig14}}
\end{figure*}

\end{document}